\documentstyle[psfig,conf_iap,10pt]{article}
\begin{document}
\def\kms{\hbox{$\rm\thinspace km s^{-1}$}}
\def\kmsMpc{\hbox{$\rm\thinspace km s^{-1} Mpc^{-1}$}}
\def\hMpc{\hbox{$\rm\thinspace h^{-1}_{75}Mpc$}}
\def\mo{\hbox{${\rm\thinspace M}_{\odot}$}}
\heading{
New Structure In The Shapley Supercluster
} 
\par\medskip\noindent
\author{
M.J. Drinkwater$^{1}$,
Q.A. Parker$^{2}$,
D. Proust$^{3}$,
H. Quintana$^{4}$,
E. Slezak$^{5}$
}
\address{%
School of Physics, University of New South Wales, Sydney 2052, Australia
}
\address{%
Anglo-Australian Observatory, Coonabarabran, NSW 2357, Australia
}
\address{%
DAEC, Observatoire de Meudon, 92195 Meudon Cedex, France
}
\address{%
Astrofisica, Universidad Cat\'olica de Chile, Casilla 104, Santiago 22, Chile
}
\address{%
Observatoire de Nice, 06304 Nice Cedex 4, France
}
\begin{abstract}
  We present new radial velocities for 189 galaxies in a 91 deg$^2$
  region of the Shapley supercluster measured with the FLAIR-II
  spectrograph on the UK Schmidt Telescope.  The data reveal two
  sheets of galaxies linking the major concentrations of the
  supercluster. The supercluster is not flattened in Declination as
  was suggested previously and it may be at least 30\% larger than
  previously thought with a correspondingly larger contribution to the
  motion of the Local Group.
\end{abstract}
\section{Wide Field Observations}

The Shapley supercluster (SSC) is recognised as one of the most
massive concentrations of galaxies in the local universe\cite{ray} so
it is of particular interest to consider its effect on the dynamics of
the Local Group\cite{pap1}. Previous studies\cite{pap1},\cite{pap2}
concentrated on the rich Abell galaxy clusters in the region, but this
might give a very biased view of the supercluster.

In this paper we present new data obtained with the FLAIR-II
multi-fibre spectrograph on the UK Schmidt Telescope at the
Anglo-Australian Observatory. This has 90 fibres in a $5.5\times5.5$
deg$^2$ field allowing us to measure a more uniform distribution of
SSC galaxies, avoiding any bias in favour of the rich clusters.  We
selected galaxies from digitised red ESO/SRC sky survey plates to a
limit of $R<16$. After removing galaxies with known redshifts and
randomly selecting between any too close to another galaxy or star to
observe we obtained samples of about 100 galaxies per Schmidt field.
We observed 3 fields centred on the SSC, obtaining velocities for a
total of 189 galaxies in the sample; full details are published
elsewhere\cite{pap3}.

The results presented in Fig.~1 show that both components of the SSC
extend much further to the South than was previously thought. These
form sheets of galaxies which extend to the full area we measured, and
presumably beyond as well. Our measurements to the North of the
cluster were much less complete (only one field in poor weather) so we
cannot exclude the possibility that these sheets of galaxies extend
equally to the North. It was earlier concluded\cite{pap1} from the
velocity distribution of the clusters that the SSC was very elongated
and either inclined towards us or rotating: we can now see that this
is not the case.
\raisebox{-2cm}[0cm][0cm]{\makebox[0cm][c]
{\hspace{2cm}\parbox{13cm}{\em Paper presented at the 14$^{th}$ IAP
Colloquium: Wide Field Surveys in Cosmology,
held in Paris, 1998 May 26--30, eds. S.Colombi, Y.Mellier
}}}

\begin{figure}
\vspace{-2.6cm}
\hbox{
\hspace{-1.4cm}
\psfig{file=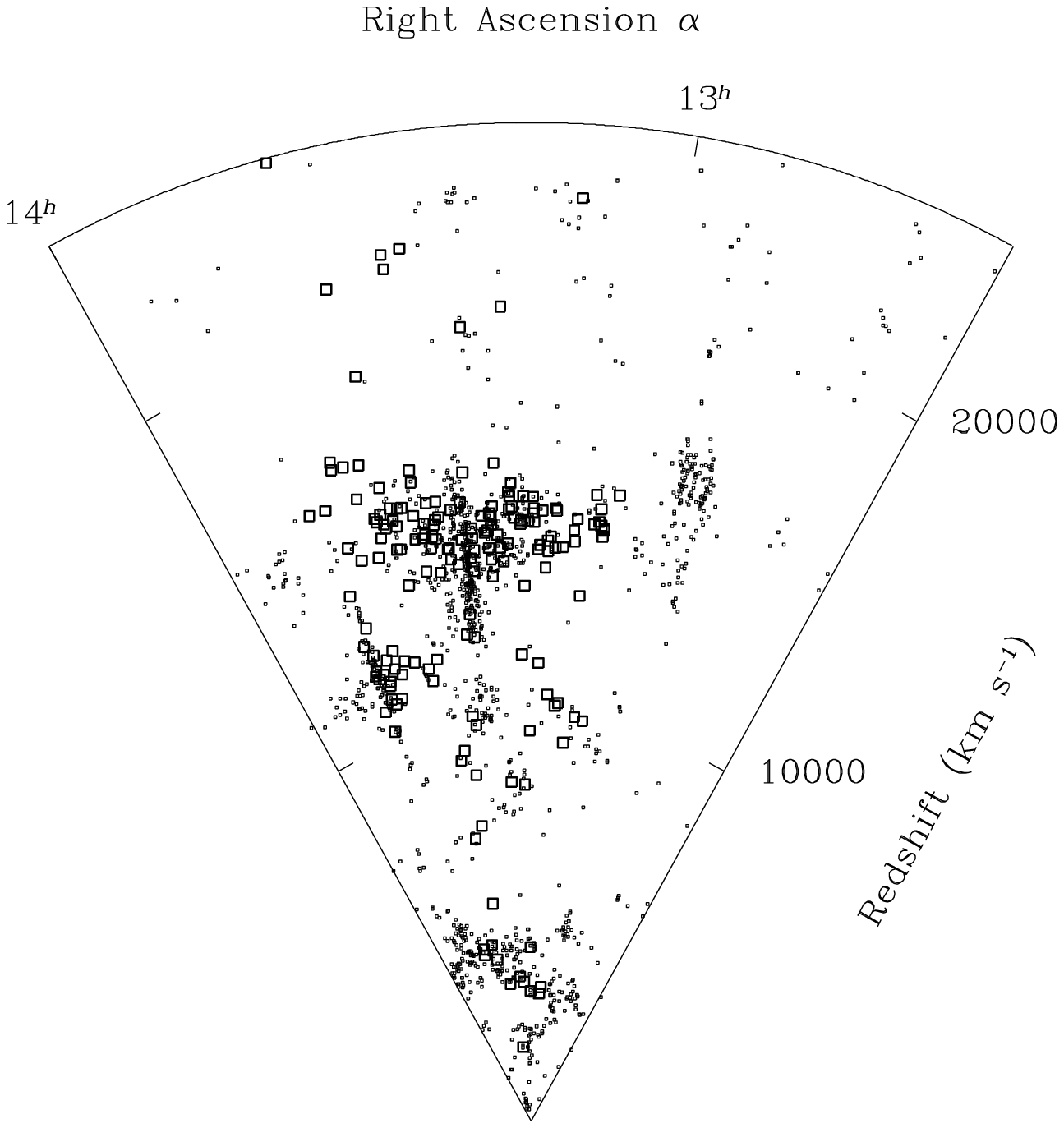,width=10.8cm}
\hspace{-5.2cm}
\psfig{file=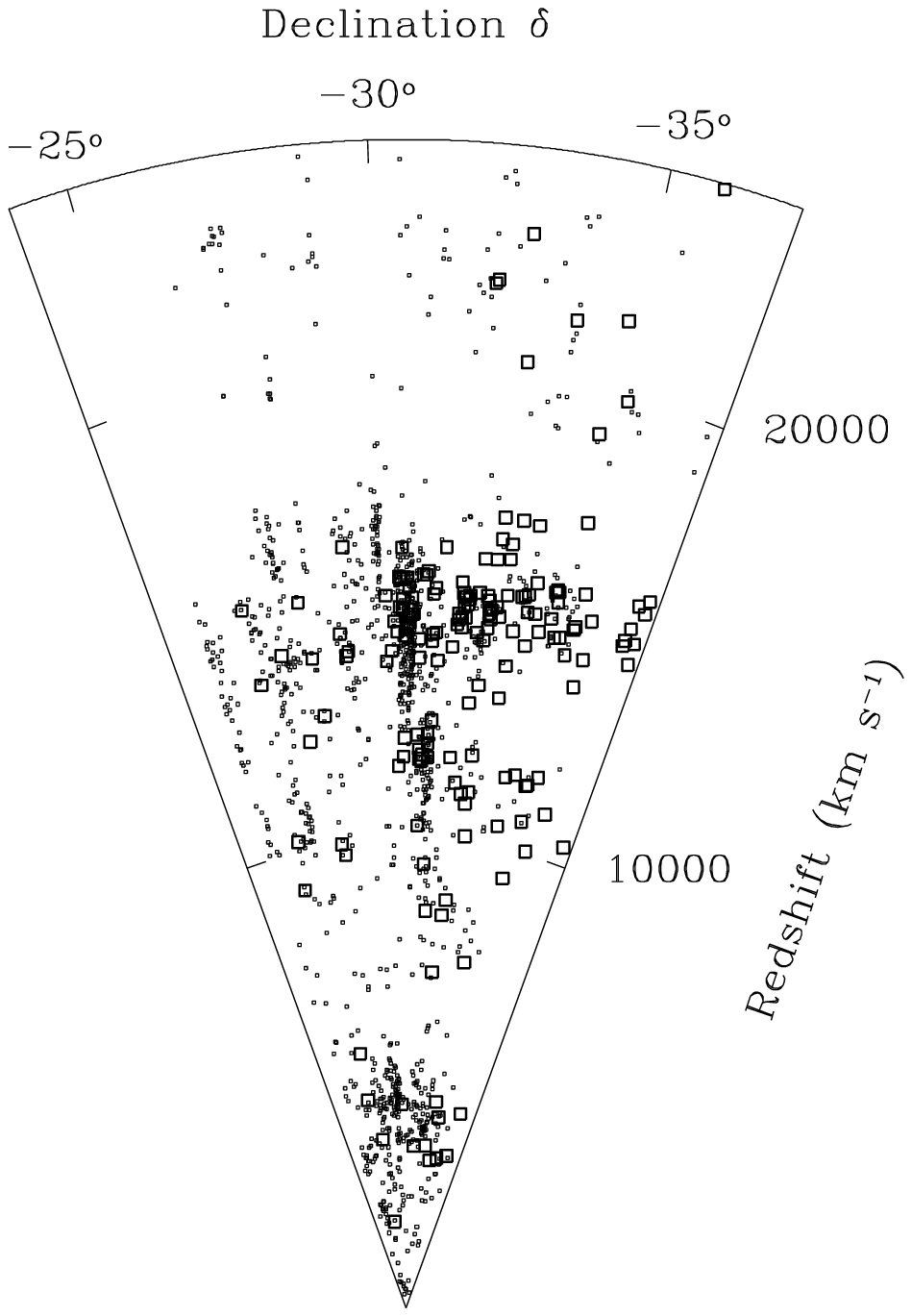,width=10.8cm}
}
\vspace{-1.2cm}
\caption[]{
  Cone diagrams of all known galaxy redshifts in the direction of the
  Shapley supercluster. Previously published galaxies are plotted as
  dots; the new measurements are plotted as open squares. We now see
  that the SSC is clearly separated into two components in velocity
  space, the nearer one at $\overline{v}=10800\kms$
  and the main concentration at $\overline{v}=14920\kms$.}
\end{figure}

We measured 152 new SSC galaxies in the velocity range
$7580<v<18300\kms$ compared to 864 previously known.  If the SSC is
equally extended to the North as to the South we might expect to find
a further additional 150 galaxies or a total of an extra 30\% of
galaxies in the central region of the SSC.  The effect of the mass of
the SSC on the dynamics of the Local Group was previously
estimated\cite{pap1} to account for at least 25\% of the motion of the
Local Group with respect to the cosmic microwave background. Our new
data suggest that the SSC is at least 30\% more massive with a
significant part of the extra mass in the closer sub-region. The SSC
therefore has an even more important effect on the Local Group than
previously thought.

\acknowledgements{This work was partly supported by the
cooperative programme ECOS/ CONICYT C96U04 and 
an ARC Special Research Initiatives grant.}

\vfill
\end{document}